\pdfoutput=1
\documentclass[fleqn,usenatbib,nofootinbib]{mnras}
\usepackage{graphicx}
\usepackage{amsmath,amssymb}
\usepackage{lastpage}
\usepackage[all]{hypcap}
\usepackage[T1]{fontenc}
\usepackage{float}
\usepackage{subfigure}
\usepackage{afterpage}
\usepackage[usenames]{color}
\usepackage{yfonts}
\usepackage{mathrsfs}
\usepackage{upgreek}
\usepackage{hyperref}
\usepackage{gensymb}

\title[Velocity and size residuals across rotation curves]{Uncorrelated velocity and size residuals across galaxy rotation curves}

\author[H. Desmond et al.]{Harry Desmond$^{1}$\thanks{E-mail: harry.desmond@physics.ox.ac.uk}, Harley Katz$^1$, Federico Lelli$^2$\thanks{ESO Fellow}, and Stacy McGaugh$^3$\\
$^1$Astrophysics, University of Oxford, Denys Wilkinson Building, Keble Road, Oxford OX1 3RH, UK\\
$^2$European Southern Observatory, Karl-Schwarzschild-Strasse 2, Garching bei Munchen, Germany\\
$^3$Case Western Reserve University, Department of Astronomy, Cleveland OH, 44106, USA\\
}

\pubyear{2018}

\begin{document}
\label{FirstPage}
\pagerange{\pageref{FirstPage}--\pageref{LastPage}}
\maketitle

\begin{abstract}
The mass--velocity--size relation of late-type galaxies decouples into independent correlations between mass and velocity (the Tully--Fisher relation), and between mass and size. This behaviour is different to early-type galaxies which lie on a Fundamental Plane. We study the coupling of the Tully--Fisher and mass--size relations in observations (the SPARC sample), empirical galaxy formation models based on halo abundance matching, and rotation curve fits with a hydrodynamically motivated halo profile. We systematically investigate the correlation coefficient between the Tully--Fisher residuals $\Delta V_r$ and mass--size residuals $\Delta R$ as a function of the radius $r$ at which the velocity is measured, and thus present the $\Delta V_r-\Delta R$ relation across rotation curves. We find no significant correlation in the data at any $r$, aside from $r \ll R_\text{eff}$ where baryonic mass dominates. We show that this implies an anticorrelation between galaxy size and halo concentration (or halo mass) at fixed baryonic mass, and provides evidence against the hypothesis that galaxy and halo specific angular momentum are proportional. Finally, we study the $\Delta V_r-\Delta R$ relations produced by the baryons and dark matter separately by fitting halo profiles to the rotation curves. The balance between these components illustrates the ``disk--halo conspiracy'' required for no overall correlation.
\end{abstract}


\section{Introduction}
\label{sec:intro}

Dynamically, to first order, a galaxy is described by a mass, $M$, a size, $R$, and a characteristic velocity, $V$. Understanding the relations between these properties as a function of cosmic time is a major goal of galaxy astrophysics, with ramifications not only for the connection between galaxies of various types and their host dark matter halos, but also for the processes that drive galaxy formation and evolution.

The $M$--$R$--$V$ relation of early-type galaxies forms a \emph{Fundamental Plane} (FP;~\citealt{Djorgovski, Dressler}), implying a single constraint between these variables. In contrast, late-type galaxies follow two separate relations, the \emph{Tully--Fisher} relation $M-V$ (TFR;~\citealt{TF}) and \emph{mass--size} relation (MSR). These are decoupled, in that their residuals $\Delta V \equiv \log(V)-\langle \log(V)|\log(M) \rangle$ and $\Delta R \equiv \log(R)-\langle \log(R)|\log(M) \rangle$ are uncorrelated (\citealt{McGaugh_res, Pizagno, Reyes, Lelli_BTFR}). While the FP may be understood as arising from virial equilibrium (for suitable choices of stellar initial mass function (IMF), dark matter fraction and radial orbit anisotropy;~\citealt{Dutton_2013, Desmond_FJR}), the independence of the TFR and MSR has been used to argue for an additional constraint between the properties of late-type galaxies that reduces the dimensionality of the effective parameter space (e.g.~\citealt{Famaey_McGaugh}).

It may be surprising that $\Delta V$ and $\Delta R$ are independent because a larger galaxy at fixed mass has a less concentrated baryonic mass profile and should therefore rotate more slowly by Kepler's laws. The baryon-only prediction $\Delta V \propto -0.5 \: \Delta R$ is amply ruled out by the data (e.g.~\citealt{McGaugh_res}). However, this neglects the effect of both the shape of the halo velocity profile (which larger galaxies sample at larger radii) and the dependence of the galaxy--halo connection on galaxy size. These effects have been modelled in different ways generating disagreement among literature studies, whose conclusions range from assertions of incompatibility between the observations and predictions of standard galaxy formation models (e.g.~\citealt{McGaugh_TTP, Lelli_BTFR}) to assertions of complete compatibility (e.g.~\citealt{Courteau_Rix, Dutton_2007}). If the independence of the TFR and the MSR is not due to an additional constraint then it must arise ``by chance'' from the interrelation of the density profiles of baryonic and dark matter: our work explores how this may come about.

There is confusion in the literature for three reasons:

\begin{enumerate}

\item{} Summarising the rotation curve (RC) of a galaxy by a single velocity, $V$, introduces a degree of arbitrariness since measuring $V$ at different radii may be expected to yield different results. Thus, while standard galaxy formation naturally predicts negligible $\Delta V-\Delta R$ correlation where RCs plateau, far beyond where most of the baryonic mass resides~\citep{Desmond_BTFR}, it may fail to do so if measured within the stellar disk (\citealt{DW15}, hereafter DW15). Different Tully--Fisher studies tend to use various definitions of radii to measure $V$ (see \citealt{Yegorova} for a comparison of different choices). The relation between these different TFRs depends on the shape of the RCs and hence on the total mass profiles.

\item{} Model $M$--$R$--$V$ relations depend crucially on the correlation between the baryonic mass profiles of late-type galaxies and the masses, $M_\text{vir}$, and concentrations, $c$, of their host haloes, which are largely responsible for setting $V$. While the $M_*-M_\text{vir},c$ relation is well known from abundance matching (AM) studies, as well as more direct observations (see~\citealt{ghc_review} and references therein), the $R-M_\text{vir},c$ relation remains mostly unconstrained. Populating more massive or more concentrated haloes with larger galaxies at fixed baryonic mass will clearly induce a positive $\Delta V-\Delta R$ correlation. Most authors impose no correlation between $R$ and halo properties at fixed $M_*$ (e.g.~\citealt{Dutton_2011, Dutton_2013, DC}), a strong assumption that neglects any potential correlation between the galaxy and halo angular momentum. Other authors impose an anticorrelation between $R$ and $c$ by assuming that the specific angular momenta of galaxies and halos are proportional (e.g. \citealt{MMW}, DW15). One can also use results from hydrodynamical simulations of galaxy formation~\citep{Desmond_EAGLE}, apply prescriptions for converting gas to stars as a function of baryonic surface density (e.g.~\citealt{Dutton_2007}), or employ tunable toy models (\citealt{Desmond_MDAR, Desmond_BTFR} and here). Given the sensitivity of the $\Delta V-\Delta R$ relation to correlation between $R$ and halo properties, it is not surprising that these studies reach apparently contradictory conclusions.

\item{} Velocity and size residuals are sensitive to the baryonic mass distributions of galaxies as well as the radii at which the RCs are sampled. It is difficult to ascertain the bias introduced by comparing model galaxies to observed ones with different mass profiles, or in cases where the mock and real observations are not made in the same way. While the use of simplistic functional forms for mass components is common, baryonic density profiles may be matched exactly between real and mock galaxies where high-quality photometry is available.

\end{enumerate}

We construct a semi-empirical $\Lambda$CDM model for galaxies in the \textit{Spitzer} Photometry and Accurate Rotation Curves (SPARC;~\citealt{SPARC}) sample by adapting and expanding previous work in~\citet{Desmond_BTFR}. Our particular interest is in comparing predicted and observed $\Delta V-\Delta R$ correlations when $V$ is measured at a range of radii across the RCs of galaxies. Besides clarifying the relative importance of various factors in setting the agreement of data and theory, we will show that this information brings new constraining power to the dependence of galaxy size on halo properties as well as the inner density profiles of haloes. We identify models in approximate agreement with the measured $\Delta V-\Delta R$ relation for all velocity choices, and hence show how the decoupling of the TFR and MSR may come about in $\Lambda$CDM.

\section{Methods}
\label{sec:method}

Our method is based on that of~\citet{Desmond_BTFR}. We begin with a sample of 153 galaxies from the SPARC database~\citep{SPARC},\footnote{\url{http://astroweb.cwru.edu/SPARC/}} requiring inclination $i\ge30$\degree and quality flag $Q<3$. We use the stellar mass of each galaxy (assuming a mass-to-light ratio of 0.5 at $3.6\mu$m for the disk and 0.7 for the bulge;~\citealt{Lelli_BTFR}) to assign a halo from the \textsc{DarkSky-400} simulation~\citep{DarkSky} by the AM prescription of~\citet{Lehmann}. We then combine the halo parameters output by \textsc{Rockstar}~\citep{Rockstar} with the measured baryon profile to create a model RC for each SPARC galaxy. We sample this RC at the same radii as the real data, and include observational error by scattering the model velocities by the quoted SPARC uncertainties. As the model is probabilistic, we adopt a Monte Carlo approach to error propagation and sample variance by analysing 200 independent realisations.

We also investigate the result of using the halo profile fits to the RCs from~\citet{Katz}. In that work, both an NFW profile, derived from $N$-body simulations, and a DC14 profile~\citep{DC14}, derived from hydrodynamical simulations, were fit to the SPARC RCs using a Markov Chain Monte Carlo (MCMC) algorithm to map out the posterior distributions of halo parameters. It was found that the DC14 profile, which can be cuspy or cored at the centre depending on $M_*/M_{\rm halo}$, could better reproduce the RCs while satisfying the stellar mass--halo mass relation from AM and the mass--concentration relation from dark matter-only simulations.

For a given radius $r$ at which $V$ is measured (see below), we calculate the velocity and radius residuals as
\begin{eqnarray}
\begin{aligned}\label{eq:1}
&\Delta V_r \equiv \log(V(r)) - \langle \log(V(r))|\log(M_\text{b}) \rangle \\
&\Delta R \equiv \log(R_\text{eff}) - \langle \log(R_\text{eff})|\log(M_\text{b}) \rangle,
\end{aligned}
\end{eqnarray}
where $M_\text{b}$ is total (cold) baryonic mass $M_* + 1.33 \: M_\text{HI}$, $R_\text{eff}$ is $3.6 \: \mu$m half-light radius, and angular brackets denote the expectation of a third-order fit to the TFR or MSR in log-space, fitting also for intrinsic scatter. The subscript $r$ highlights the dependence of $V$ on the radius at which it is measured. We then calculate $\rho_\text{sr}(r)$ as the Spearman's rank coefficient of the $\Delta V_r-\Delta R$ correlation, and repeat the analysis for each model realisation.

To measure the strength of the $\Delta V_r-\Delta R$ correlation across the RCs, we calculate $V$ at either $r = x R_\text{eff}$ or $r = x R_\text{p}$, where $x$ is a universal variable in the range [$0.2, 6$] and $R_\text{p}$ is the radius at which the baryonic component of the RC peaks~\citep{McGaugh_res}. As the median $R_\text{eff}$ across the sample is $3.1$ kpc and the median $R_\text{p}$ is $5.4$ kpc, this probes the RCs in the range $\sim0.6-30$ kpc. We remove galaxies with RCs that do not extend to $x R_\text{eff}$ or $x R_\text{p}$, which is $\sim50\%$ of the sample for $x < 0.4$ and $\sim70\%$ for  $x > 5$. The use of $R_\text{eff}$ versus $R_\text{p}$ corresponds to different relative $r$ values between the galaxies at fixed $x$, according to their baryon mass distributions. In each case we calculate the median $\rho_\text{sr}(r)$ values across the model realisations as well as the standard deviation among them.

We compare the data to a series of models of increasing complexity. In our fiducial model, the halo profile is unaffected by galaxy formation, AM is described by the best-fit parameters of~\citet{Lehmann} ($\alpha_\text{AM}=0.6$, $\sigma_\text{AM}=0.16$ dex) and galaxy size is uncorrelated with halo properties at fixed $M_\text{b}$. By varying these assumptions we exhibit the sensitivity of the $\rho_\text{sr}-r$ relation to them; we will show in particular that this relation brings important new constraining power to the relation between galaxy size and halo mass or concentration.

\section{Results}
\label{sec:results}

\begin{figure*}
\centerline{\includegraphics[]{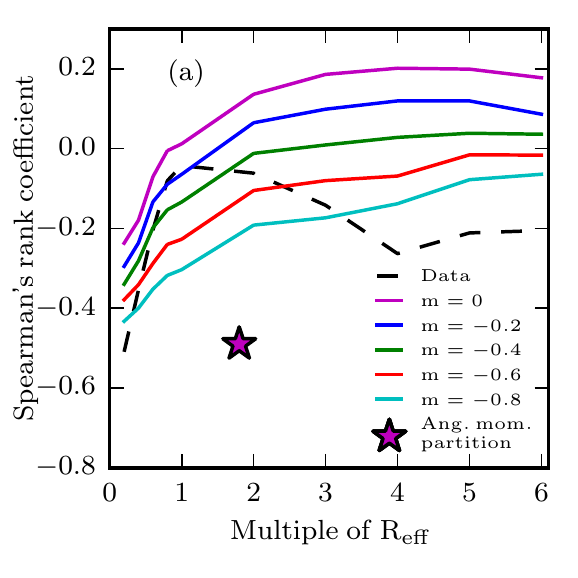}\includegraphics[]{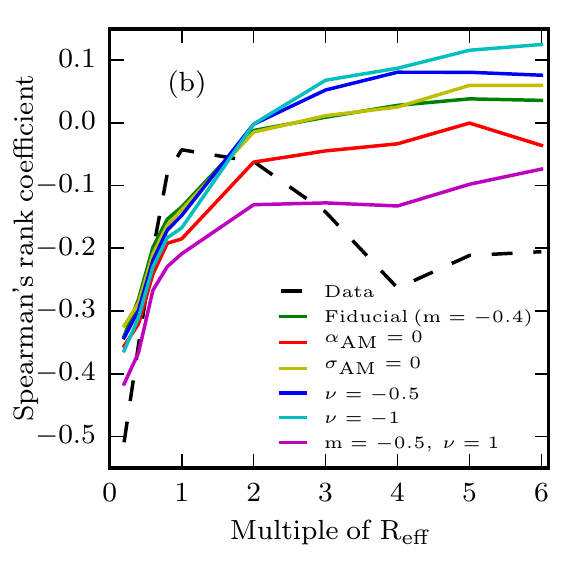}\includegraphics[]{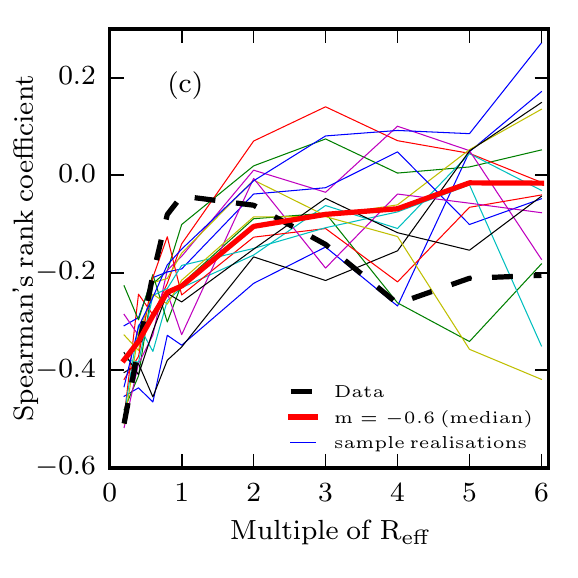}}
  \caption{Spearman's rank coefficient $\rho_\text{sr}$ of the $\Delta V_r-\Delta R$ correlation as a function of the multiple of $R_\text{eff}$ at which the velocity is measured. We show the results both in the actual SPARC data (black dashed line), and the median of $200$ SPARC-like realisations drawn from various theoretical models (coloured lines left and centre). In panel (a), the models vary in the strength of anticorrelation between galaxy size and halo concentration at fixed baryonic mass, as quantified by $m$ in $\Delta c = m \Delta R_\text{eff}$. The data favours a moderate anticorrelation. The purple star indicates the result of a model in which galaxy and halo specific angular momentum are proportional~\citep{DW15}. In panel (b) other parameters are varied individually around the $m=-0.4$ model, including the AM proxy $\alpha_\text{AM}$ and scatter $\sigma_\text{AM}$ and the response of the halo profile to galaxy formation. $\nu=1$ describes standard adiabatic expansion~\citep{Gnedin_2011}, while $\nu<0$ indicates halo expansion. In panel (c) we show the results of 15 individual realisations of the $m=-0.6$ model of panel (a) to illustrate the sample variance. Given this variance the SPARC $\Delta V_r-\Delta R$ relation is not unlikely, indicating this model to be satisfactory (see also Fig.~\ref{fig:2}). In no case is the $\Delta V_r-\Delta R$ relation statistically significant.}
  \label{fig:1}
\end{figure*}

We begin with $V(xR_\text{eff})$. As the dashed black line of Fig.~\ref{fig:1}(a) we show $\rho_\text{sr}(r)$ for the SPARC galaxies as $x$ is varied. We find the |$\rho_\text{sr}$| values to be fairly small for all $r$, indicating that $\Delta V_r$ and $\Delta R$ are always insignificantly correlated, although there is a moderate anticorrelation for $r \ll R_\text{eff}$. The prediction of the fiducial model is depicted as the solid magenta line, which also shows a statistically insignificant correlation but with a $\rho_\text{sr}$ value somewhat larger than in the data. The dependence of $\rho_\text{rs}$ on $r$ in the model is set by a combination of two factors:

\begin{enumerate}

\item{} As $r$ decreases, the baryonic contribution to $V$ increases and hence the more $V$ is reduced when the galaxy is made larger at fixed $M_\text{b}$. This induces a negative $\rho_\text{sr}$. This is the dominant effect for $r \lesssim R_\text{eff}$ ($x \lesssim 1$) where the greater part of $V$ is set by the baryons. At very small radii ($r\lesssim0.2\:R_\text{eff}$) $\rho_\text{sr}$ tends to the baryon-only result, which we show below to be $\sim-0.6$.

\item{} Since the radius at which the velocity is measured scales with the size of a galaxy, smaller galaxies sample the halo RC at smaller radius. If halo properties do not depend on $R_\text{eff}$ at fixed $M_\text{b}$ -- as is the case in our fiducial model -- this simply reduces $V$ if the halo RC is rising, inducing a positive $\rho_\text{sr}$. This is the dominant effect at large $x$ where the halo RC is still rising but the baryonic contribution to $V$ is small. At very large $x$ (beyond the halo scale radius) the halo RC starts to decrease, and hence $\rho_\text{sr}$ does also. This is difficult to trace out because few SPARC galaxies have velocity measurements at such large galactocentric radii, although there is a slight indication of it in the model for $x \gtrsim 5$.

\end{enumerate}

The other lines in Fig.~\ref{fig:1}(a) show the result of introducing a correlation between the residuals of galaxy size and halo concentration, of the form $\Delta c = m \Delta R_\text{eff}$ with 0.1 dex scatter in $c$ at fixed $M_\text{b}$ (cf. Eq.~\ref{eq:1};~\citealt{Desmond_MDAR, Desmond_BTFR}). For $m<0$ this puts larger galaxies in less concentrated halos at fixed $M_\text{b}$, causing a reduction in $\Delta V_r$ as $\Delta R$ is increased. Overall agreement with the data is maximised for $-0.8 \lesssim m \lesssim -0.4$, which agrees well with the constraint of~\citet{Desmond_MDAR}, where $m$ was inferred from the correlation of residuals of the mass discrepancy--acceleration (or radial acceleration) relation with galaxy size, and of~\citet{Desmond_BTFR} which examined the $\Delta V-\Delta R$ correlation with $V$ measured at $R_\text{flat}$. Although $m<0$ could be inferred using any radius choice within the range we consider, the full $\Delta V-\Delta R$ relation demonstrates that a moderate anticorrelation of $\Delta R_\text{eff}$ with $\Delta c$ improves agreement with the observations regardless of the choice of radius. Note also that halo mass could have been used rather than concentration: the relevant quantity is the dark matter mass within $r$, which is a function of both. We use concentration because it has a larger scatter than $M_\text{vir}$ at fixed $M_\text{b}$ due to a weaker correlation with the AM proxy.

Finally, the purple star in Fig.~\ref{fig:1}(a) shows the result of setting galaxy and halo specific angular momentum proportional to one another (\citealt{MMW}), specifically from the model of DW15 who find $\langle \rho_\text{sr} \rangle=-0.49$ when $V$ is measured at $R_{80} \simeq 1.8 \: R_\text{eff}$. This model is strongly disfavoured by the data; we discuss this result further in Section~\ref{sec:discussion}.

In Fig.~\ref{fig:1}(b), we show the results of varying other model parameters around the $m=-0.4$ line of Fig.~\ref{fig:1}(a), including the AM proxy $\alpha_\text{AM}$ and scatter $\sigma_\text{AM}$, and the response of the halo to galaxy formation. The latter is quantified by $\nu$ (\citealt{Dutton_2007}; DW15; \citealt{Desmond_FJR}): $\nu>0$ corresponds to contraction from a primordial NFW form, and $\nu<0$ to expansion. If $\nu > 0$ the halo velocity profile rises more steeply in the inner regions and less steeply further out. This causes $\rho_\text{sr}$ to rise at small $x$ and fall at large $x$, and vice versa if the halo expands. However, this effect is small and largely degenerate with $m$. The best fit to the data is given by the \textit{a priori} plausible case $m=-0.5$, $\nu=1$ (purple). Adiabatic contraction ($\nu=1$) is however disfavoured by other kinematical measurements (\citealt{Dutton_2007}; DW15; \citealt{Desmond_FJR}), so that a more likely solution is a lower $\nu$ and slightly lower $m$.

\begin{figure}
  \centering
  \includegraphics[scale=1.]{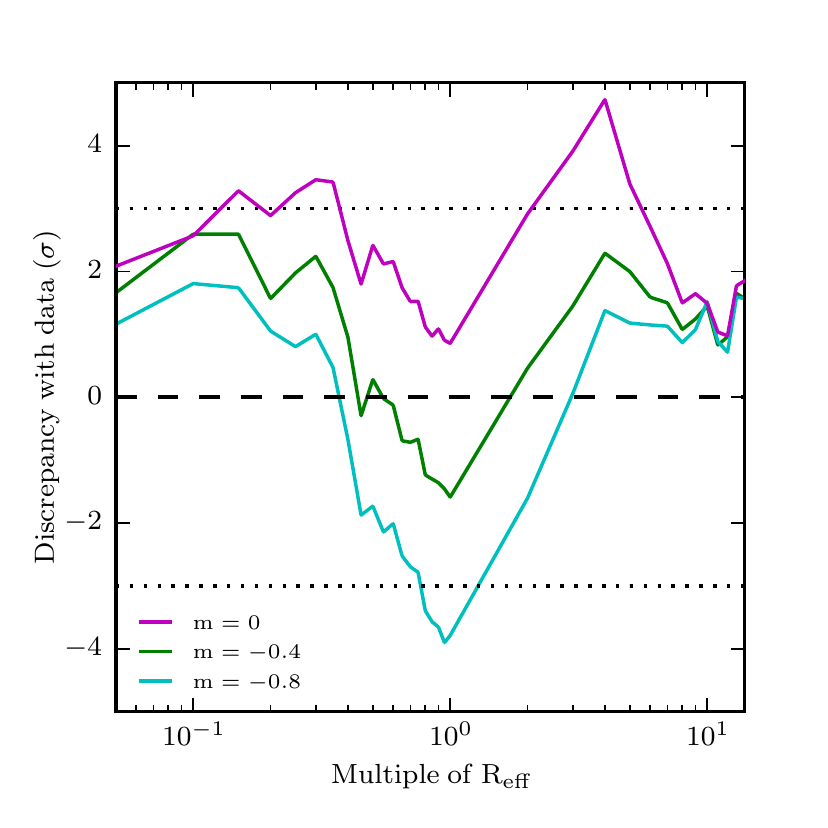}
  \caption{Significance of the model-data discrepancies $\delta_\text{sr}$ (Eq.~\ref{eq:discrepancy}) as a function of the multiple of $R_\text{eff}$ at which $V$ is measured, accounting for sample variance among model realisations. We use a logarithmic $x$-axis to expand the $r<R_\text{eff}$ region, and show dotted lines at $\pm3\sigma$ to assess statistical significance. While models with no correlation between $R_\text{eff}$ and halo concentration or mass at fixed $M_\text{b}$ significantly overpredict $\rho_\text{sr}$ across the RCs (magenta line; cf. Fig.~\ref{fig:1}a) and those with a strong anticorrelation often underpredict it (cyan), a moderate anticorrelation yields a $\delta_\text{sr}<3\sigma$ discrepancy for all $r$ (green).}
  \label{fig:2}
\end{figure}

Figs.~\ref{fig:1}(a) and~\ref{fig:1}(b) show the average $\rho_\text{sr}$ from many realisations of the model. However, to assess the compatibility of the $\Delta V_r-\Delta R$ measurements and model predictions one must ask whether the measurements could plausibly have been drawn from the model, which requires knowledge of the sample variance among mock data sets. In Fig.~\ref{fig:1}(c) we show the results for 15 randomly chosen individual realisations of the model with $m=-0.6$. This spread is sufficient to render the data not unlikely were the model true. We quantify this in Fig.~\ref{fig:2}, where we plot the discrepancy between the measured $\rho_\text{sr}$ values and the expectations from various models as a function of $x$. We define the discrepancy as
\begin{equation}
\delta_\text{sr} \equiv (\rho_\text{sr,obs} - \langle \rho_\text{sr} \rangle)/\sigma(\rho_\text{sr}), \label{eq:discrepancy}
\end{equation}
where $\rho_\text{sr,obs}$ is the value from the SPARC data, $\langle \rangle$ denotes the median average of the model realisations and $\sigma$ the standard deviation between them. The fiducial model in which $R_\text{eff}$ is independent of halo properties at fixed $M_\text{b}$ has a maximum $\delta_\text{sr} = 4.7\sigma$ discrepancy with the data at $x \approx 4$, indicating statistically significant evidence for an anticorrelation of $\Delta R_\text{eff}$ and $\Delta c$ or $\Delta M_\text{vir}$. The model with $\Delta c = -0.4\:\Delta R_\text{eff}$ has a maximum discrepancy of $\delta_\text{sr} = 2.6\sigma$. In principle even stronger evidence for $m<0$ could be acquired by combining the correlated information across the full range of $x$.

\begin{figure}
  \centering
  \includegraphics[scale=1.]{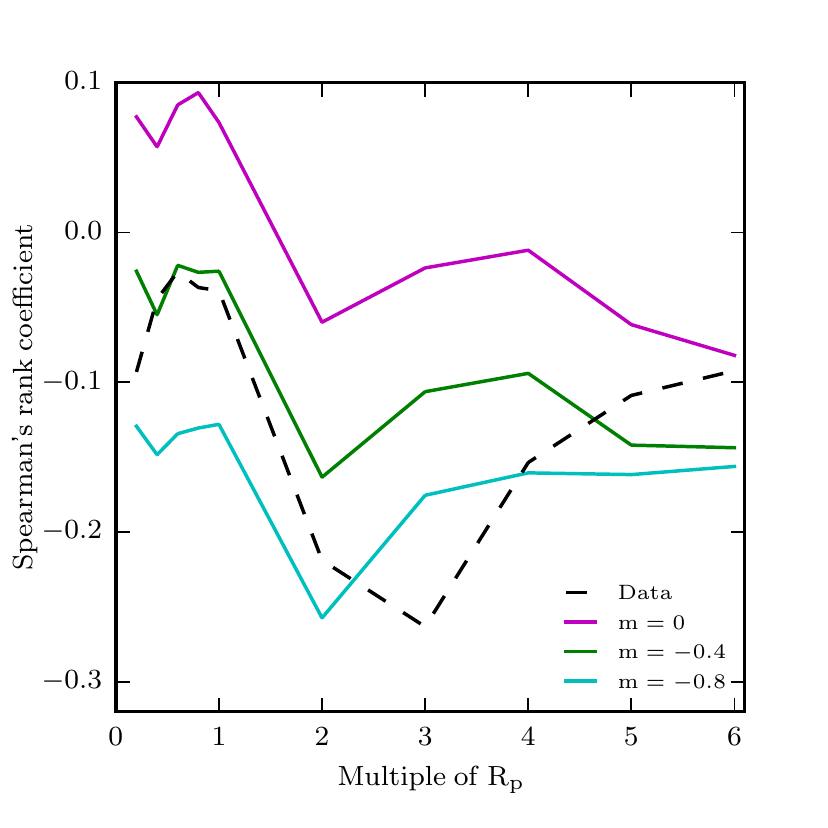}
  \caption{As Fig.~\ref{fig:1}, except with $V$ measured at a multiple of the radius $R_\text{p}$ at which the baryonic RC peaks, rather than $R_\text{eff}$. Again $\Delta V_r$ and $\Delta R$ are not significantly correlated in either data or model, and there is evidence for a $\Delta R_\text{eff}-\Delta c$ anticorrelation.}
  \label{fig:3}
\end{figure}

In Fig.~\ref{fig:3} we show the analogue of Fig.~\ref{fig:1} when $V$ is measured at a multiple of the radius $R_\text{p}$ at which the baryonic RC peaks, rather than $R_\text{eff}$. Although the dependence of $\rho_\text{sr}$ on $r$ is different owing to variations among the galaxies in the radii relative to $R_\text{eff}$ at which $R_\text{p}$ is achieved, the best models of Fig.~\ref{fig:1} again provide a good fit to the data. This demonstrates that our results are not very sensitive to the particular choice of normalisation for the radius definition.

We have checked that neither our data nor model relations are sensitive to outliers by excising galaxies lying $>$2$\sigma$ from the fitted $M_\text{b}-R_\text{eff}$ and $M_\text{b}-V(x R_\text{eff})$ relations. We find that $\rho_\text{sr}$ changes by at most $\sim$0.05, and the outlier fraction is $<$10\%. We have also checked that bootstrap and jackknife resamples of the SPARC data set have qualitatively similar $\rho_\text{sr}-r$ relations.

\begin{figure*}
\centerline{\includegraphics[scale=1.]{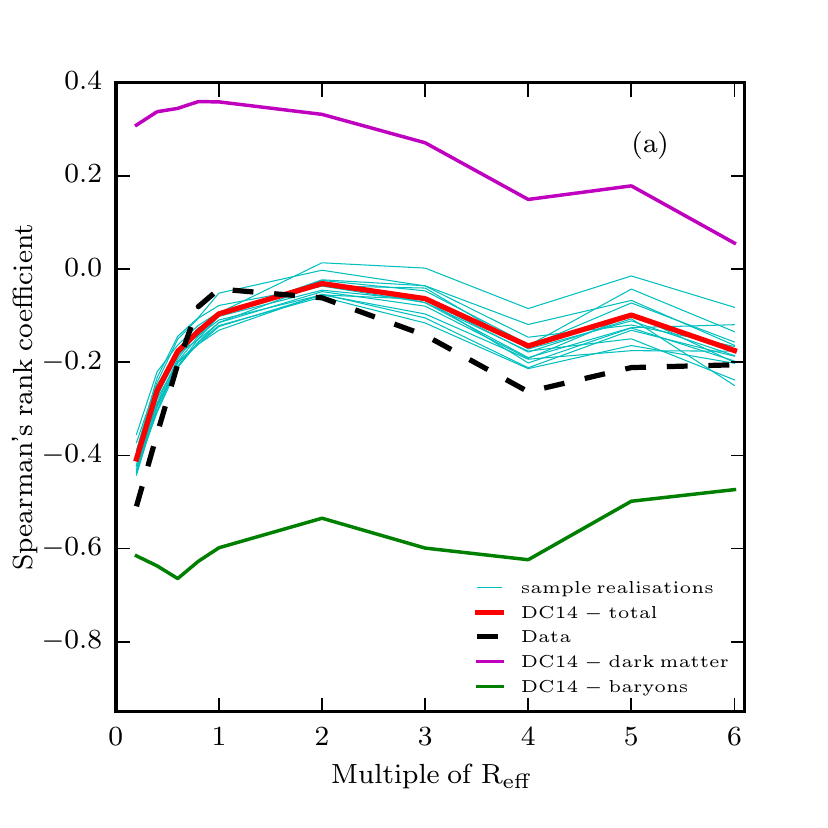}\includegraphics[scale=1.]{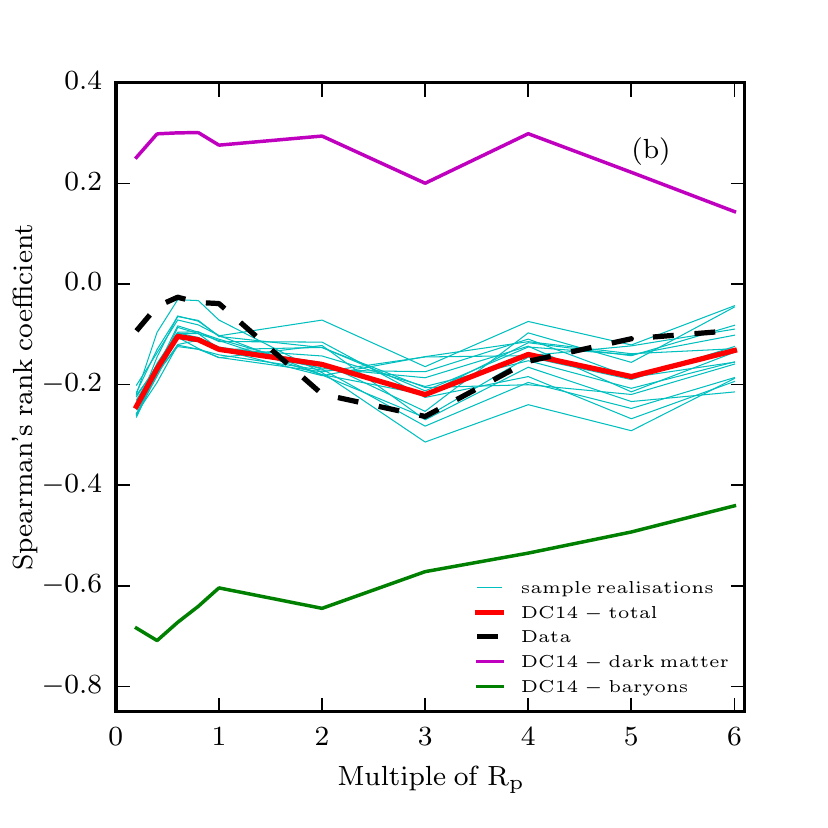}}
  \caption{As Figs.~\ref{fig:1} and \ref{fig:3}, but for the DC14 halo profile fit to the SPARC RCs. Panel (a) is for $r=xR_\text{eff}$ and panel (b) is for $r=xR_\text{p}$. The dashed black lines show the observations, the red lines the median total RCs from the halo model over 200 Monte Carlo realisations, the thin cyan lines 15 example realisations and the green and magenta lines the medians using only the baryonic and dark matter parts of the RC respectively. While the baryons alone would produce a significantly negative $\Delta V_r-\Delta R$ correlation in each case, and the halo a moderately positive correlation, the combination produces the very weak and largely scale-invariant anticorrelation found in the data.}
  \label{fig:4}
\end{figure*}

Finally, in Fig.~\ref{fig:4} we show the results from fitting the RCs with the DC14 profile, using either $V(xR_\text{eff})$ (Fig.~\ref{fig:4}a) or $V(xR_\text{p})$ (Fig.~\ref{fig:4}b). The red line shows the median over 200 model realisations drawn from the joint posterior on $M_*/L$ and host halo mass and concentration for each galaxy \citep{Katz}.  The thin lines show 15 example realisations.  The model matches the $\Delta V_r-\Delta R$ relation well, with only a small range of $r$ lying outside the model realisations. Note however that this is not a forward model but uses information from the RCs themselves, and is therefore tuned to match the data: the red and dashed black lines are not independent. Fig.~\ref{fig:4} simply shows that halo profiles known to reproduce some aspects of galaxy kinematics are also able to decouple the Tully--Fisher and mass--size relations.

We also show in that figure the results using the RCs generated by only the baryons (green) or dark matter (magenta). As expected, the baryons provide a strongly negative contribution to $\rho_\text{sr}$: a larger galaxy at fixed $M_\text{b}$ produces a smaller rotation velocity because its baryonic mass is more diffuse. The halo however provides a positive $\rho_\text{sr}$, since larger galaxies sample the halo RC at larger $r$ for fixed $x$ and $M_\text{b}$. This effect is less pronounced at larger $r$ where the halo RC is less steeply rising: far enough out the halo RC would begin to decline, causing $\rho_\text{sr} < 0$. The combination of these components in the total RC produces the $|\rho_\text{sr}|$ values near 0 that are measured. This provides a new perspective on the ``disk--halo conspiracy''~\citep{conspiracy}: not only must the relative distributions of baryons and dark matter conspire to make RCs flat, they must also decouple the sizes of galaxies from their velocities across the RCs.

\section{Discussion}
\label{sec:discussion}

We have shown that realistic models for the galaxy--halo connection produce statistically insignificant correlations between the residuals of the Tully--Fisher and mass--size relations. These models illustrate the ``conspiracy'' between the distributions of baryons and dark matter required to decouple the characteristic velocities and sizes of galaxies, no matter where those velocities are measured. This does not require fine tuning but rather holds for a range of model assumptions. Here we discuss our result in the context of related studies in the literature.

In DW15, $V$ was measured at the radius enclosing $80$\% of the $i$-band light, which is $1.8 \: R_\text{eff}$ for an exponential disk. The DW15 model gives a significantly stronger $\Delta V_r-\Delta R$ anticorrelation than any of the models we investigate here ($\rho_\text{sr} \simeq -0.5$; see their fig. 6). The reason for this is that DW15 assumed proportionality between the specific angular momentum of baryons and dark matter to set galaxy size, which strongly anticorrelates $\Delta R$ with halo concentration at fixed galaxy mass (i.e. implies a strongly negative $m$;~\citealt{Desmond_EAGLE}). As we show here, $\Delta V_r$ and $\Delta R$ are never as strongly anticorrelated as predicted by that model outside of the baryon-dominated region $r \ll R_\text{eff}$, and the predicted $\rho_\text{sr}$ decreases further at smaller $r$. This provides further evidence that galaxy size is not set entirely by equipartition of specific angular momentum between baryonic and dark mass. Note that DW15 assumed baryonic mass models that differ in detail from those of the SPARC galaxies, making their results not fully commensurable with the present data. However, observational studies agree on the weakness of the $\Delta V-\Delta R$ correlation~\citep{McGaugh_res,Pizagno,Reyes,Lelli_BTFR}, so it is unlikely that the measured $\rho_\text{sr}$ depend sensitively on details of the SPARC mass models. DW15 modelled self-consistently the correlations of all halo properties from the \textsc{DarkSky} simulation, including the anticorrelation between concentration and spin at fixed mass~\citep{Maccio}. The conclusion that setting $R_\text{eff}$ proportional to halo spin significantly anticorrelates $\Delta R$ and $\Delta V$ is in agreement with similar models~\citep{Dutton_2007}. Nevertheless, these works do not exclude the possibility of a successful model for galaxy size based on halo spin, but only show that it cannot take the simplest form of a direct proportionality between baryonic and dark matter specific angular momentum when making standard assumptions for halo density profiles and other aspects of the galaxy--halo connection (e.g. AM).

Another early study of the $\Delta V_r-\Delta R$ relation was~\citet{Courteau_Rix}, which used the lack of observed anticorrelation with $V$ measured at $2.2$ disk scale lengths to infer the relative amount of dark and visible matter within that radius. Our study is more general in that we do not restrict ourselves to a single radius, and more precise in that we tailor our models to the observational data set in question. We are therefore able to generalise the conclusion that prior-motivated halo models produce dark matter fractions consistent with those required to generate a negligible $\Delta V_r-\Delta R$ correlation, and provide more detailed information on the conditions for maximal agreement with the data.

We have found evidence for an anticorrelation of $R_\text{eff}$ with $c$ (or $M_\text{halo}$) at fixed $M_\text{b}$. This is also found in~\citet{Desmond_MDAR,Desmond_BTFR} and produced in the EAGLE hydrodynamical simulation~\citep{Desmond_EAGLE}. We note however that $M_\text{vir}$ is likely \emph{positively} correlated with $\Delta R$: not only is this produced in simulations such as EAGLE~\citep{Desmond_EAGLE}, it is also measured with weak lensing~\citep{charlton}. The correlation that we infer may be understandable in the future through more detailed physical modelling of the relation between galaxy and halo angular momentum.

Our model for the $\Delta R_\text{eff}-\Delta c$ correlation complements other methods in the literature for incorporating size into the galaxy--halo connection. One alternative is to impose a proportionality between galaxy and halo size, which reproduces the shape of the mass--size relation~\citep{Kravtsov_sizes} as well as the size dependence of galaxy clustering~\citep{Hearin_sizes}. This approach is motivated by the angular momentum partition model of~\citet{MMW}, although formally independent of it. It is not yet clear what $R_\text{eff} \propto R_\text{vir}$ implies for the relation between $R_\text{eff}$ and halo properties at fixed stellar or baryonic mass, i.e. after factoring out the principal component of the galaxy--halo connection $M_\text{*/b}-M_\text{vir},c$. A third method for connecting size to halo properties is conditional abundance matching~\citep{CAM} where a second AM is performed on size in bins of stellar mass. Although this naturally models size at fixed galaxy mass, making it orthogonal to $M_*-M_\text{vir},c$, and reproduces by construction the size function conditioned on $M_*$, it is unclear whether it can account for the kind of dynamical signals investigated here. Future work should aim to integrate these approaches into a single unified model for the role of size in the galaxy--halo connection and draw out the implications for angular momentum transfer.

\section{Conclusions}
\label{sec:conc}

Using $153$ late-type galaxies from the SPARC sample, we investigate the correlation between the residuals of the mass--size and baryonic Tully--Fisher relations with velocities measured at a range of radii. Our main findings are the following:

\begin{itemize}

\item{} The correlation between the velocities and sizes of galaxies at fixed baryonic mass is a weak function of the radius $r$ at which the velocity is measured, with Spearman's rank coefficient $\rho_\text{sr}$ rising from $\sim-0.5$ in the baryon-dominated inner regions ($r \ll R_\text{eff}$) to $\sim-0.2$ further out where the dark matter is more important. The full radial dependence of this relation provides new information about the dependence of galaxy size on halo properties.

\item{} Models that set $M_*$ by abundance matching and assume that $R_\text{eff}$ is uncorrelated with halo properties at fixed galaxy mass overpredict the strength of the $\Delta V_r-\Delta R$ relation by $1-5\sigma$, depending on $r$. This suggests an anticorrelation of galaxy size with halo concentration (or mass) at fixed baryonic mass, in line with previous inferences from the $M$--$R$--$V$ relations of late-type galaxies. We show agreement within $3\sigma$ for all $r$ using a model in which $\Delta c \simeq -0.4\:\Delta R_\text{eff}$.

\item{} The $\rho_\text{sr}-r$ relation may also be matched by fitting the RCs with a partly cored DC14 halo profile. We show explicitly the $\Delta V_r-\Delta R$ correlation produced by the dark matter and anticorrelation produced by the baryons, thus quantifying the ``baryon--halo conspiracy'' required for no overall correlation at any $r \gtrsim R_\text{eff}$.

\item{} The $\rho_\text{sr}-r$ relation provides further evidence against the hypothesis that galaxy and halo specific angular momentum are proportional. We conclude that, under standard assumptions for halo density profiles and the galaxy--halo connection, this putative proportionality cannot be responsible for setting galaxy size at low redshift.

\end{itemize}

\section*{Acknowledgements}

HD is supported by St John's College, Oxford.  HK thanks the Beecroft Fellowship, Brasenose College, and the Nicholas Kurti Junior Fellowship.

\bsp

\end{document}